\documentclass[11pt]{article}

\usepackage{fullpage}

\usepackage{amsmath}
\usepackage{amssymb}
\usepackage{graphicx}
\newcommand{\subf}[1]{%
  {\small\begin{tabular}[t]{@{}c@{}}
  #1
  \end{tabular}}%
}
\newcommand{\url}[1]{{\tt #1}}

\usepackage[ruled,vlined,linesnumbered]{algorithm2e}

\renewcommand{\log}{\lg}

\newcommand{\rank}{\mathit{rank}}

\newcommand{\select}{\mathit{select}}

\newcommand{\excess}{\mathit{excess}}
\newcommand{\raiz}{\mathit{root}}
\newcommand{\fchild}{\mathit{fchild}}
\newcommand{\lchild}{\mathit{lchild}}
\newcommand{\nsibling}{\mathit{nsibling}}
\newcommand{\psibling}{\mathit{psibling}}
\newcommand{\parent}{\mathit{parent}}
\newcommand{\isleaf}{\mathit{isleaf}}

\newcommand{\degree}{\mathit{degree}}

\newcommand{\depth}{\mathit{depth}}
\newcommand{\height}{\mathit{height}}
\newcommand{\deepestnode}{\mathit{deepestnode}}
\newcommand{\subtree}{\mathit{subtree}}
\newcommand{\isancestor}{\mathit{isancestor}}
\newcommand{\levelanc}{\mathit{levelancestor}}
\newcommand{\preorder}{\mathit{preorder}}
\newcommand{\postorder}{\mathit{postorder}}

\newcommand{\preorderselect}{\mathit{preorderselect}}
\newcommand{\postorderselect}{\mathit{postorderselect}}

\newcommand{\children}{\mathit{children}}
\newcommand{\child}{\mathit{child}}
\newcommand{\childrank}{\mathit{childrank}}

\newcommand{\lca}{\mathit{lca}}

\newcommand{\numleaves}{\mathit{numleaves}}
\newcommand{\leafrank}{\mathit{leafrank}}
\newcommand{\leafselect}{\mathit{leafselect}}
\newcommand{\leftmostleaf}{\mathit{leftmostleaf}}
\newcommand{\rightmostleaf}{\mathit{rightmostleaf}}

\newcommand{\levelnext}{\mathit{levelnext}}
\newcommand{\levelprev}{\mathit{levelprev}}
\newcommand{\levelleftmost}{\mathit{levelleftmost}}
\newcommand{\levelrightmost}{\mathit{levelrightmost}}

\newcommand{\fwdsearch}{\mathit{fwdsearch}}
\newcommand{\bwdsearch}{\mathit{bwdsearch}}
\newcommand{\close}{\mathit{close}}
\newcommand{\open}{\mathit{open}}
\newcommand{\enclose}{\mathit{enclose}}
\newcommand{\rmq}{\mathit{rmq}}
\newcommand{\rMq}{\mathit{rMq}}
\newcommand{\mincount}{\mathit{mincount}}
\newcommand{\minselect}{\mathit{minselect}}

\newcommand{\Oh}[1]{\mathcal{O}\!\left(#1\right)}
\newcommand{\polylog}{\mathop{\mathrm{polylog}}}

\DontPrintSemicolon
\SetKwInOut{Input}{\footnotesize Input}
\SetKwInOut{Output}{\footnotesize Output}
\SetKw{KwAnd}{and}
\SetKw{KwOr}{or}
\SetKw{KwNot}{not}
\SetKw{KwTrue}{true}
\SetKw{KwFalse}{false}

\SetKwProg{Fn}{Proc}{}{}

\begin{document}

\title{Simple and Efficient Fully-Functional Succinct Trees}

\author{Joshimar Cordova and Gonzalo Navarro \\
Department of Computer Science, University of Chile \\
{\tt \{jcordova|gnavarro\}@dcc.uchile.cl}}

\maketitle 

\begin{abstract}
The fully-functional succinct tree representation of Navarro and Sadakane
({\em ACM Transactions on Algorithms}, 2014) supports a large number of
operations in constant time using $2n+o(n)$ bits. 
However, the full idea is hard to implement. Only a simplified version with
$\Oh{\log n}$ operation time has been implemented and shown to be practical 
and competitive. We describe a new variant of the original idea that is much 
simpler to implement and has worst-case time $\Oh{\log\log n}$ for the 
operations. An implementation based on this version is experimentally shown to
be superior to existing implementations.
\end{abstract}

\section{Introduction}

Combinatorial arguments show that it is possible to represent any ordinal
tree of $n$ nodes using less than $2n$ bits of space: the number of such trees
is the $(n-1)$th Catalan number, $\frac{1}{n}{2n-2 \choose n-1}$, and its
logarithm (in base 2 and written $\lg$ across this paper) is $2n-\Theta(\log n)$. A simple way to
encode any ordinal tree in $2n$ bits is the so-called {\em balanced
parentheses (BP)} representation: traverse the tree in depth-first order,
writing an opening parenthesis upon reaching a node, and a closing one upon
definitely leaving it. Much more challenging is, however, to efficiently
navigate the tree using that representation.

The interest in navigating a $2n$-bit representation of a tree, compared to
using a classical $\Oh{n}$-pointers representation, is that those succinct
data structures allow fitting much larger datasets in the faster and
smaller levels of the memory hierarchy, thereby improving the overall system
performance. Note that compression is not sufficient; it must be possible to
operate the data in its compressed form. The succinct representation of 
ordinal trees is one of the most clear success stories in this field. 
Table~\ref{tab:ops} lists the operations that can be supported in constant 
time within $2n+o(n)$ bits of space. These form a rich set that suffices for
most applications.

\begin{table}[t!]
\caption{Operations on ordinal trees, where $i$ and $j$ are node identifiers.}
\label{tab:ops}
\footnotesize
\begin{center}
  \begin{tabular}{lll|l}
  operation & description \\ \hline
%  $\inspect(i)$ & $P[i]$ \\
%  $\close(i)$ / $\open(i)$ & position of parenthesis matching $P[i]$ \\
%  $\enclose(i)$ & position of tightest open parenthesis enclosing $i$ \\
%  $\rankopen(i)$ / $\rankclose(i)$ & number of open/close parentheses in $P[0,i]$ \\
%  $\selectopen(k)$ / $\selectclose(k)$ & position of $k$th open/close parenthesis \\
%  $\rmq(i,j)$ / $\rMq(i,j)$ & position of min/max excess value in range $[i,j]$ \\
%\hline
  $\raiz$ & the tree root \\
  $\preorder(i)$ / $\postorder(i)$ & preorder/postorder rank of node $i$ \\
  $\preorderselect(k)$ / $\postorderselect(k)$ & the node with preorder/postorder $k$ \\
  $\isleaf(i)$ & whether the node is a leaf \\
  $\isancestor(i,j)$ & whether $i$ is an ancestor of $j$ \\
  $\depth(i)$ & depth of node $i$ \\
  $\parent(i)$ & parent of node $i$ \\
  $\fchild(i)$ / $\lchild(i)$ & first/last child of node $i$ \\
  $\nsibling(i)$ / $\psibling(i)$ & next/previous sibling of node $i$ \\
  $\subtree(i)$ & number of nodes in the subtree of node $i$ \\
  $\levelanc(i,d)$ & ancestor $j$ of $i$ such that $\depth(j) = \depth(i)-d$ \\
  $\levelnext(i)$ / $\levelprev(i)$ & next/previous node of $i$ with the same depth \\
  $\levelleftmost(d)$ / $\levelrightmost(d)$ & leftmost/rightmost node with depth $d$ \\
  $\lca(i,j)$ & the lowest common ancestor of two nodes $i,j$ \\
  $\deepestnode(i)$ & the (first) deepest node in the subtree of $i$ \\
  $\height(i)$ & the height of $i$ (distance to its deepest node) \\
  $\degree(i)$ & $q=$ number of children of node $i$ \\
  $\child(i,q)$ & $q$-th child of node $i$ \\
  $\childrank(i)$ & $q=$ number of siblings to the left of node $i$ \\
%  $\inorder(i)$ & inorder of node $i$ \\
%  $\inorderselect(i)$ & node with inorder $i$ \\
  $\leafrank(i)$ & number of leaves to the left and up to node $i$ \\
  $\leafselect(k)$ & $k$th leaf of the tree \\
  $\numleaves(i)$ & number of leaves in the subtree of node $i$ \\
  $\leftmostleaf(i)$ / $\rightmostleaf(i)$ & leftmost/rightmost leaf of node $i$ \\
  \hline
  \end{tabular}
\end{center}
\end{table}

The story starts with Jacobson \cite{Jac89}, 
who proposed a simple levelwise representation called LOUDS, which reduced 
tree navigation to two simple primitives on bitvectors: $\rank$ and 
$\select$ (all these primitives will be defined later). 
However, the repertoire of tree operations was limited. Munro and Raman 
\cite{MR01} used for the first time the BP representation, and showed how 
three basic primitives on the parentheses: $\open$, $\close$, and $\enclose$, 
plus $\rank$ and $\select$, were sufficient to support a significantly wider 
set of operations. The operations were supported in constant time, however
the solution was quite complex in practice. Geary et al.~\cite{GRRR06} retained 
constant times with a much simpler solution to $\open$, $\close$, and 
$\enclose$, based on a two-level recursion scheme. Still, not all the
operations of Table~\ref{tab:ops} were supported. Missing ones were
added one by one: $\children$ \cite{CLL05}, $\levelanc$ \cite{MRRR12}, 
$\child$, $\childrank$, $\height$, and $\lca$ \cite{LY08}. Each such addition
involved extra $o(n)$-bit substructures that were also hard to 
implement.

An alternative to BP, called DFUDS, was introduced by Benoit et 
al.~\cite{BDMRRR05}. It also used $2n$ balanced parentheses, but they had a
different interpretation. Its main merit was to support $\child$ and related 
operations very easily and in constant time. It did not support, however, 
operations $\childrank$, $\depth$, $\levelanc$, and $\lca$, which were added 
later \cite{GRR06,JSS12}, again each using $o(n)$ bits and requiring a complex
implementation to achieve constant time.

Navarro and Sadakane \cite{NS14} introduced a new representation based on BP,
said to be {\em fully-functional} because it supported all of the operations
in Table~\ref{tab:ops} in constant time and using a single set of structures.
This was a significant simplification of previous results and enabled the
development of an efficient implementation. The idea was to reduce all the
tree operations to a small set of primitives over parentheses:
$\fwdsearch$, $\bwdsearch$, $\rmq$, and a few variations. The main structure
to implement those primitives was the so-called {\em range min-max tree
(rmM-tree)}, which is a balanced tree of arity $\log^\epsilon n$ (for a constant
$0<\epsilon<1$) that supports the primitives in constant time on buckets of 
$\Oh{\polylog n}$ parentheses. To handle queries that were not solved 
within a bucket, other structures had to be added, and these were far less simple.

A simple $\Oh{\log n}$-time implementation using a single binary range min-max 
tree for the whole sequence \cite{ACNS10} was shown to be faster (or use much less space, or both)
than other implementable constant-time representations \cite{GRRR06} in several
real-life trees and navigation schemes. Only the LOUDS representation was
shown to be competitive, within its limited functionality. 
While the $\Oh{\log n}$ growth was shown to be imperceptible in many real-life
traversals, some stress tests pursued later \cite{JR12} showed
that it does show up in certain plausible situations.

No attempt was made to implement the actual constant-time proposal \cite{NS14}.
The reason is that, while constant-time and $o(n)$-bit space in theory, the 
structures used for inter-bucket queries, as well as the variant of rmM-trees 
that operates in constant time, involve large constants and include structures
that are known to be hard to implement efficiently, such as fusion 
trees \cite{FW93} and compressed bitvectors with optimal redundancy 
\cite{Pat08}. Any practical
implementation of these ideas leads again to the $\Oh{\log n}$ times already
obtained with binary rmM-trees.

In this paper we introduce an alternative construction that builds on binary
rmM-trees and is simple to implement. It does not reach constant times, but
rather $\Oh{\log\log n}$ time, and requires $2n+\Oh{n/\log n}$ bits of space.
We describe a new implementation building on these ideas, and experimentally 
show that it outperforms a state-of-the-art implementation of the 
$O(\log n)$-time solution, both in time and space, and therefore becomes the
new state-of-the art implementation of fully-functional succinct trees.

\section{Basic Concepts}

\subsection{Bits and balanced parentheses}

Given a bitvector $B[1,2n]$, we define $\rank_t(i)$ as the number of 
occurrences of the bit $t$ in $B[1,i]$. We also define $\select_t(k)$ as
the position in $B$ of the $k$th occurrence of the bit $t$. Both primitives
can be implemented in constant time using $o(n)$ bits on top of $B$ 
\cite{Cla96}. Note that $\rank_1(i)+\rank_0(i)=i$ and $\rank_t(\select_t(k))=k$.

A sequence of $2n$ parentheses will be represented as a bitvector $B[1,2n]$
by interpreting `(' as a 1 and `)' as a 0. On such a sequence we define the
operation $\excess(i)$ as the number of opening minus closing parentheses in
$B[1,i]$, that is, $\excess(i) = \rank_1(i)-\rank_0(i) = 2\rank_1(i)-i$.
We say that $B$ is {\em balanced} if $\excess(i) \ge 0$ for all $i$, and 
$\excess(2n)=0$. Note that $\excess(i)=\excess(i-1)\pm 1$.

In a balanced sequence, every opening parenthesis at $B[i]$ has a matching 
closing parenthesis at $B[j]$ for $j>i$, and every other parenthesis opening
inside $B[i+1,j-1]$ has its matching parenthesis inside $B[i+1,j-1]$ as well.
Thus the parentheses define a hierarchy. Moreover, we have 
$\excess(j)=\excess(i)-1$ and $\excess(m) \ge \excess(i)$ for all $i < m < j$. 
This motivates the definition of the following primitives on parentheses
\cite{MR01}:
\begin{description}
        \item[$\close(i)$:] the position of the closing parenthesis that matches
		$B[i]=1$, that is, the smallest $j>i$ such that
                $\excess(j)=\excess(i)-1$.
        \item[$\open(i)$:] the position of the opening parenthesis that matches
		$B[i]=0$, that is, the largest $j<i$ such that
                $\excess(j-1)=\excess(i)$.
        \item[$\enclose(i)$:] the opening parenthesis of the smallest matching 
		pair that contains position $i$, that is, the largest $j<i$
		such that $\excess(j-1)=excess(i)-2$.
\end{description}

It turns out that a more general set of primitives is useful to implement a
large number of tree operations \cite{NS14}, which look forward or backward
for an arbitrary relative excess:
\begin{eqnarray*}
\fwdsearch(i,d)
	&\!=\!& \min \{ j>i,~\excess(j)=excess(i)+d \}, \\
\bwdsearch(i,d)
	&\!=\!& \max \{ j<i,~\excess(j)=excess(i)+d \}.
\end{eqnarray*}
In particular, we have $\close(i)=\fwdsearch(i,-1)$, 
$\open(i)=\bwdsearch(i,0)+1$, and $\enclose(i)=\bwdsearch(i,-2)+1$.

To implement other tree operations, we also need the following primitives,
which refer to minimum and maximum excess in a range of $B$:
\begin{description}
\item[$\rmq(i,j)$:] position of the leftmost minimum in $\excess(i)$,
$\excess(i+1),\ldots,\excess(j)$.
\item[$\rMq(i,j)$:] position of the leftmost maximum in $\excess(i)$,
$\excess(i+1),\ldots,\excess(j)$.
\item[$\mincount(i,j)$:] number of occurrences of the minimum in $\excess(i)$,
$\excess(i+1),\ldots,\excess(j)$.
\item[$\minselect(i,j,q)$:] position of the $q$th minimum in $\excess(i)$,
$\excess(i+1),\ldots,\excess(j)$.
\end{description}

\subsection{BP representation of ordinal trees}

\begin{figure}[t]
\centerline{\includegraphics[width=0.8\textwidth]{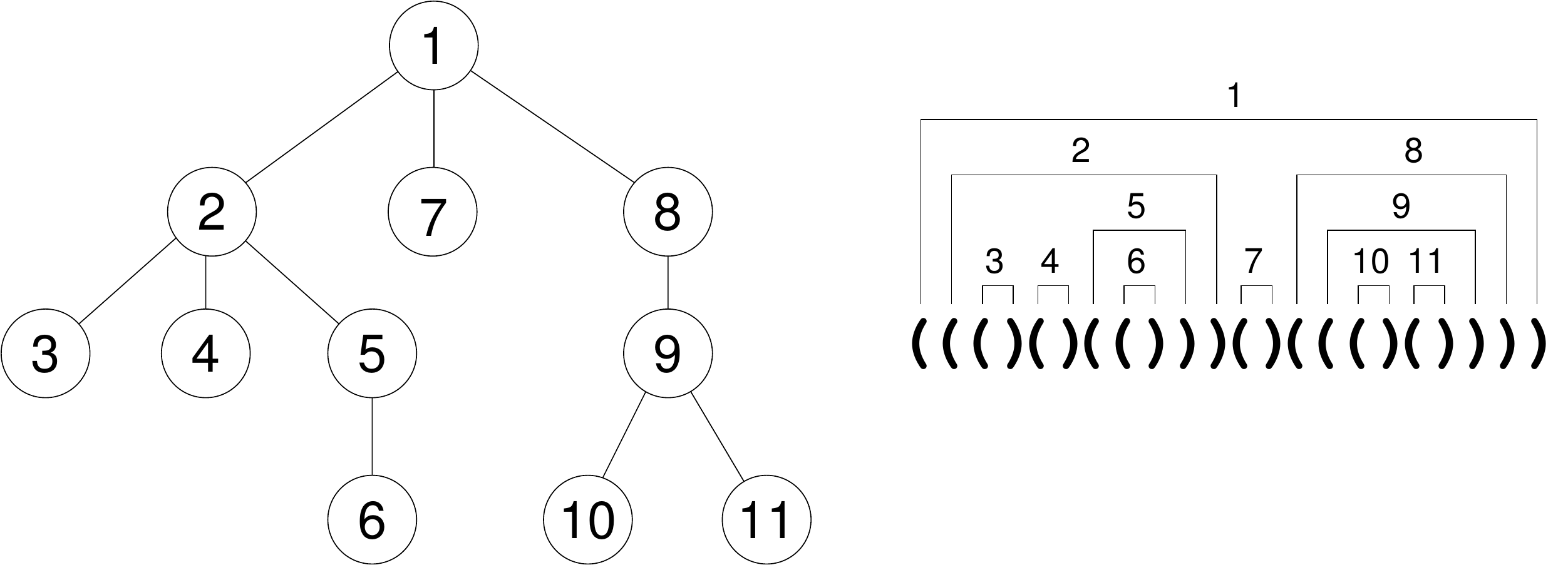}}
\caption{An ordinal tree on the left (the node identifiers are their preorder
numbers) and its BP representation on the right, indicating which parentheses
represent each node. For example the node with preorder 5 has identifier 7,
which is its position in the sequence of parentheses.}
\label{fig:tree}
\end{figure}

As said in the Introduction, an ordinal tree of $n$ nodes is represented with
$2n$ parentheses by opening a parenthesis when we arrive at a node and closing
it when we leave the node. The resulting sequence is balanced, and the
hierarchy it defines corresponds to subtree containment. Let us identify
each node with the position of its opening parenthesis in the sequence $B$.
See Figure~\ref{fig:tree}.

Many tree operations are immediately translated into the primitives 
we have defined \cite{MR01}: $\raiz=1$,
$\depth(i)=\excess(i)$, $\parent(i)=\enclose(i)$,
$\isleaf(i)$ iff $B[i+1]=0$, $\fchild(i)=i+1$ (if $i$ is not a leaf),
$\nsibling(i)=\close(i)+1$ (if the result $j$ holds $B[j]=0$ then $i$ has no
next sibling), $\psibling(i)=\open(i-1)$ (if $B[i-1]=1$ then $i$ has no
previous sibling), $\lchild(i) = \open(\close(i)-1)$ (if $i$ is not a leaf),
$\preorder(i)=\rank_1(i)$, $\preorderselect(k)=\select_1(k)$,
$\postorder(i)=\rank_0(\close(i))$, $\postorderselect(k)=\open(\select_0(k))$,
$\isancestor(i,j)$ iff $i\le j <\close(i)$, and $\subtree(i)=(\close(i)-i+1)/2$.

The primitives $\fwdsearch$ and $\bwdsearch$ yield other tree 
operations \cite{NS14}: $\levelanc(i,d)=\bwdsearch(i,-d-1)+1$,
$\levelnext(i)=\fwdsearch(\close(i),1)$,
$\levelprev(i)=\open(\bwdsearch(i,0)+1)$,
$\levelleftmost(d)=\fwdsearch(0,d)$, and
$\levelrightmost(d)=\open(\bwdsearch(2n+1,d))$.
The other primitives yield the remaining operations:
$\degree(i)=\mincount(i+1,\close(i)-1)$,
$\child(i,q)=\minselect(i+1,\close(i)-1,q-1)+1$ for $q>1$
(for $q=1$ it is $\fchild(i)$),
$\childrank(i)=\mincount(\parent(i)+1,i)+1$ unless $B[i-1]=1$
(in which case $\childrank(i)=1$),
$\lca(i,j) = \parent(\rmq(i,j)+1)$ unless $\isancestor(i,j)$
(so $\lca(i,j)=i$) or $\isancestor(j,i)$ (so $\lca(i,j)=j$),
$\deepestnode(i)=\rMq(i,\close(i))$, and
$\height(i)=\excess(\deepestnode(i))-\excess(i)$.

Finally, the operations on leaves are solved by extending the bitvector 
$\rank$ and $\select$ primitives to count the occurrences of pairs $10$ (which 
represent tree leaves, `()'): $\rank_{10}(i)$ is the number of occurrences of 
$10$ starting in $B[1,i]$ and $\select_{10}(k)$ is the position of the $k$th
occurrence of $10$ in $B$. These are easily implemented as extensions of the 
basic $\rank$ and $\select$ primitives, adding other $o(n)$ bits on top of
$B$. Then 
$\leafrank(i)=\rank_{10}(i)$, $\leafselect(k)= \select_{10}(k)$, 
$\numleaves(i)=\leafrank(\close(i))-\leafrank(i-1)$,
$\leftmostleaf(i) = \leafselect(\leafrank(i-1)+1)$ and finally
$\rightmostleaf(i) = \leafselect(\leafrank(\close(i)))$.

Therefore, all the operations of Table~\ref{tab:ops} are supported via the
primitives $\fwdsearch$, $\bwdsearch$, $\rmq$, $\rMq$, $\mincount$, and
$\minselect$. We also need $\rank$ and $\select$ on 0, 1, and 10.

\subsection{Range min-max trees}

\begin{figure}[t]
\centerline{\includegraphics[width=0.8\textwidth]{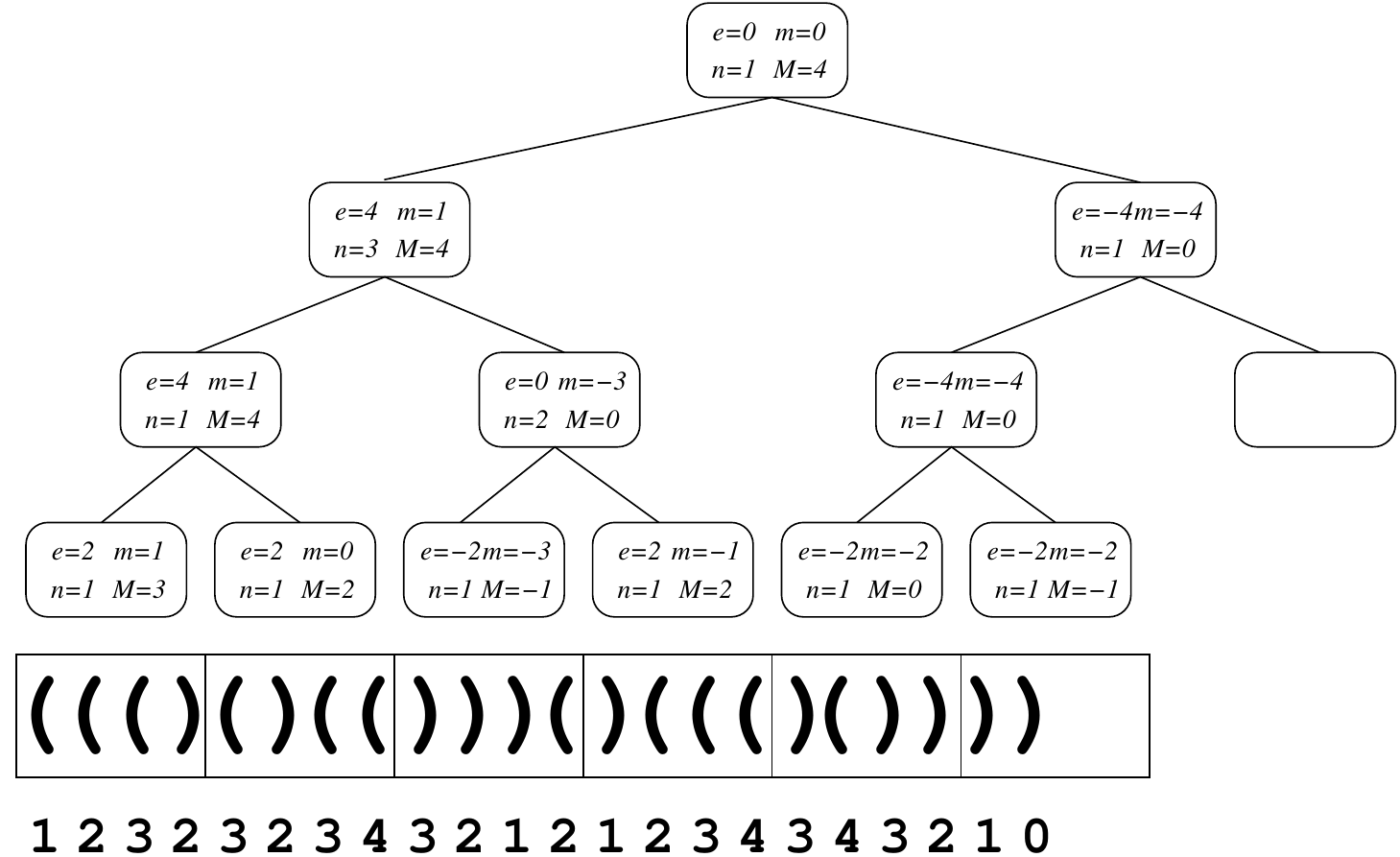}}
\caption{The rmM-tree of the parentheses sequence of Figure~\ref{fig:tree}.
The numbers below are $\excess(i)$.}
\label{fig:rmM}
\end{figure}

We describe the simple version of the structure used by Navarro and Sadakane 
\cite{NS14} to solve the primitives. 
We choose a block size $b$. Then, the (binary) range 
min-max tree, or rmM-tree, of $B[1,2n]$ is a complete binary tree where the 
$k$th leaf covers $B[(k-1)b+1,kb]$. Each rmM-tree node $v$ stores the following
fields: $v.e$ is the total excess of the area covered by $v$, $v.m$ is the
minimum excess in this area, $v.M$ is the maximum excess in the area, and
$v.n$ is the number of times the minimum excess occurs in the area. Since the
rmM-tree is complete, it can be stored without pointers, like a heap.
See Figure~\ref{fig:rmM}.

Then, an operation like $\fwdsearch(i,d)$ is solved as follows. First, the block
number $k=\lceil i/b\rceil$ is scanned from position $i+1$ onwards, looking for 
the desired excess. If not found, then we reset the desired relative excess
to $d \leftarrow d - (\excess(kb)-\excess(i))$ and consider the leaf $v$ of 
the rmM-tree that covers block $k$. Now we move upwards from $v$, looking for 
its nearest ancestor that contains the answer. At every step, if $v$ is a
right child, we move to its parent. If it is a left child, we see if
$v'.m \le d \le v'.M$, where $v'$ is the (right) sibling of $v$. If $d$ is not 
in the range, then update $d \leftarrow d - v'.e$ and move to the parent of $v$.
At some point in the search, we find that $v'.m \le d \le v'.M$ for the
sibling $v'$ of $v$, and then start descending from $v'$. Let $v_l$ and $v_r$ 
be its left and right children, respectively. 
If $v_l.m \le d \le v_l.M$, then we descend to $v_l$. Otherwise,
we update $d \leftarrow d - v_l.e$ and descend to $v_r$. Finally, we arrive at
a leaf, and scan its block until finding the excess $d$.
Operation $\bwdsearch(i,d)$ is analogous; we scan in the other direction.

For $\rmq(i,j)$, we scan the blocks of $i$ and $j$ and, if there are blocks
in between, we consider the fields $v.m$ of the $\Oh{\log \frac{j-i}{b}}$
maximal nodes that cover the leaves contained in $B[i,j]$. Then we identify
the minimum excess in $B[i,j]$ as the minimum found across the scans and the
maximal nodes. If the first occurrence of the minimum is inside the scanned blocks, that
position is $\rmq(i,j)$. Otherwise, we must start from the node $v$ that
contained the first occurrence of the minimum and traverse downwards, looking
if the first occurrence was to the left or to the right (by comparing the
fields $v.m$). Operation $\rMq(i,j)$ is analogous. For $\mincount(i,j)$ we
retraverse the blocks and nodes, adding up the fields
$v.n$ of the nodes where $v.m$ is the minimum. Finally,
for $\minselect(i,j,q)$, we do the same counting but traverse downward from
the node $v$ where the $q$th occurrence is found, to find its position.

Finally, for primitives $\rank_t(i)$ and $\select_t(k)$, we
can compute on the fly the number of 1s inside any node $v$ as
$v.r = (|v|+v.e)/2$, where $|v|$ is the size of the area of $B$ covered by
$v$. For $\rank_1(i)$, we count the 1s
in the block of $i$ and then climb upwards from the leaf $v$ covering $i$,
adding up $v'.r$ for each left sibling of $v$ found towards the root. For
$\rank_0(i)$ we compute $i-\rank_1(i)$. For $\select_1(k)$, we start from the
root node $v$, going to the left child $v_l$ if $v_l.r \ge k$, and otherwise
updating $k \leftarrow k-v_l.r$ and going to the right child. For
$\select_0(k)$ we proceed analogously, but using $|v_l|-v_l.r$ instead of
$v_l.r$. Finally, $\rank$ and $\select$ on $10$ is implemented analogously,
but we need to store a field $v.rr$ storing the number of 10s.

By using small precomputed tables that allow us to scan any block of 
$c=(\log n)/2$ bits in constant time (i.e., computing the analogous to fields
$e$, $m$, $M$, and $n$ for any chunk of $c$ bits), the total time of the
operations is $\Oh{b/c + \log n}$ bits. The extra space of the rmM-tree over 
the $2n$ bits of $B$ is $\Oh{(n/b)\log n}$ bits. For example, we can use a
single rmM-tree for the whole $B$, set $b=\log^2 n$, and thus have all the
operations implemented in time $\Oh{\log n}$ within $2n+\Oh{n/\log n}$ bits.
This is essentially the practical solution implemented for this structure
\cite{ACNS10}.

\section{An $\Oh{\log\log n}$ Time Solution} 

Now we show how to obtain $\Oh{\log\log n}$ worst-case time, still within
$\Oh{n/\log n}$ extra bits. The main idea (still borrowing from the original
solution \cite{NS14}) is to cut $B[1,2n]$ into $n'=2n/\beta$ buckets of 
$\beta = \Theta(\log^3 n)$ bits. We maintain one (binary) rmM-tree for each 
bucket. The block size of the rmM-trees is set to $b=\log n \log\log n$. 
This maintains the extra space of each rmM-tree within
$\Oh{(\beta/b)\log\beta}$ bits, adding up to $\Oh{(n/b)\log\beta}=\Oh{n/\log n}$
bits. Their operation times also stay $\Oh{b/c+\log\beta}=\Oh{\log\log n}$.

Therefore, the operations that are solved within a bucket take $\Oh{\log\log n}$
time.  The difficult part is how to handle the operations that span more than 
one bucket: a $\fwdsearch(i,d)$ or $\bwdsearch(i,d)$ whose answer is not found
within the bucket of $i$, or a $\rmq(i,j)$ or similar operation where $i$
and $j$ are in different buckets.

For each bucket $k$, we will store an entry $e[k]=\excess(k\beta)$ with the
excess at its end, and entries $m[k]=\min_{(k-1)\beta <i \le k\beta}
\excess(i)$ and $M[k]=\max_{(k-1)\beta <i \le k\beta}
\excess(i)$ with the minimum and maximum absolute excess reached inside the
bucket. These entries require just $\Oh{n'\log n}=\Oh{n/\log^2 n}$ bits of 
space. Heavier structures will be added for each operation, as described next.

\subsection{Forward and backward searching}  \label{sec:fwd-large}

The solution for these queries is similar to the original one \cite{NS14},
but we can simplify it and make it more practical by allowing us to take
$\Oh{\log\log n}$ time to solve the operation. 
We describe its details for completeness.

We first try to solve $\fwdsearch(i,d)$ inside the bucket of $i$, 
$k^*=\lceil i/\beta \rceil$. If the answer is found in there, we have completed
the query in $\Oh{\log\log n}$ time. Otherwise, after scanning the block, we
have computed the new relative excess sought $d$ (which is the original one
minus $\excess(k^*\beta)-\excess(i)$). This is converted into absolute with
$d \leftarrow d + e[k^*]$. 

Now we have to find the answer in the buckets $k^*+1$ onwards. We have to find
the smallest $k>k^*$ with $m[k] \le d \le M[k]$, and then find the answer inside
bucket $k$. Let us first consider the next bucket. If 
$m[k^*+1] \le d \le M[k^*+1]$, then the desired excess is reached inside the 
next bucket, and therefore we complete the query by running 
$\fwdsearch(0,d-e[k^*])$ inside the rmM-tree of bucket $k^*+1$.
Otherwise, either $d < m[k^*+1]$ or $d > M[k^*+1]$. Let us consider the first
case, as the other is symmetric (and requires other similar data structures).
The query $\bwdsearch(i,d)$ works
similarly, except that we look towards the left, therefore it is also analogous.

Since the excess changes by $\pm 1$ from one parenthesis to the next, 
it must hold $M[k+1] \ge m[k]-1$ for all
$k$, that is, there are no holes in the ranges $[m[k],M[k]]$ of consecutive
buckets. Therefore, if $d < m[k^*+1]$, then we simply look for the smallest
$k>k^*+1$ such that $m[k] \le d$.
Note that for this search we would like to consider, given a $k$ where
$m[k] > d$, only the smallest $k' > k$ such that $m[k'] < m[k]$, as those
values $m[k+1],\ldots,m[k'-1] \ge m[k]$ are not the solution. If we define
a tree where $k'$ is the parent of $k$, then we are looking for the nearest
ancestor $k''$ of node $k$ where $m[k''] < d$.

The solution builds on a well-known problem called {\em level-ancestor queries}
(an operation we have already considered for our succinct trees). Given a node
$v$ and a distance $t$, we want the ancestor at distance $t$ from $v$. In the
classical scenario, there is an elegant and simple solution to this problem
\cite{BF04}.
It requires $\Oh{n'\log^2 n'}$ bits of space, but this is just $\Oh{n/\log n}$.
The idea is to extract the longest root-to-leaf path and write it on an array
called a {\em ladder}. Extracting this path disconnects the tree into several 
subtrees. Each disconnected subtree is processed recursively, except that each 
time we write a path $p[1,\ell]$ of nodes into a new ladder, we continue 
writing the ancestors up to other $\ell$ nodes. That is, a path $p[1,\ell]$ 
is converted into a ladder of $2\ell$ nodes (or less if we reach the global
root). Thus the ladders add up to at most $2n'$ cells.

In the ladders,
each node has a {\em primary} copy, corresponding to the path $p[1,\ell]$ where
it belongs, and zero or more {\em secondary} copies, corresponding to paths
that are extended in other ladders. We store a pointer to the primary copy of
each node, and the id of its ancestors at distances $t=2^l$, for $l=0,1,\ldots$.
This is where the $n'\log n'$ words of space are used.

Now, to find the $t$th ancestor of $d$, we compute $l=\lfloor \log t \rfloor$,
and find in the tables the ancestor $u$ at distance $2^l$ of $v$. Then we go
to the ladder where the primary copy of $u$ is written. Because we
extract the longest paths, since $u$ has height at least $2^l$, the path
$p[1,\ell]$ where it belongs must be of length at least $2^l$, and therefore
the ladder is of length at least $2\ell \ge 2\cdot 2^l$. Therefore, the ladder
contains the ancestors of $u$ up to distance at least $2^l$, and thus the one
we want, at distance $t-2^l < 2^l$, is written in the ladder. Thus we just read
the answer in that ladder and finish.

We must extend this solution so that we find the first ancestor $u$ with 
$m[u] \le d$. Recall that the values $m[u]$ form a decreasing sequence as we
move higher in the sequence of ancestors, and within any ladder. First, we
can find the appropriate $l$ value with a binary search in the ancestors at
distance $2^l$, so that $l$ is the smallest one such that the ancestor $u$
at distance $2^l$ still has $m[u] > d$. This takes $\Oh{\log\log n'}$ time.

Now, in the ladder of $u$, we must find the first cell $u'$ to its
right with $m[u'] \le d$. We solve this by representing all the $m[u']$ values
as $B[m[u']]=1$ in a bitvector $B$ created for that ladder. Then 
$\rank_1(B,d)$ is the distance from the end of the ladder to the position of
the desired ancestor $u'$.

A useful bitvector representation for this matter is the \emph{sarray} by
Okanohara and Sadakane \cite[Sec.~6]{OS07}.\footnote{Other compressed
representations use $o(u)$ further bits, which make them unsuitable for us.}
If the ladder contains $r$
elements and the maximum value is $\mu$, then it takes $r\log\frac{\mu}{r} +
\Oh{r}$ bits of space (which adds up to just $\Oh{n'\log n}$ bits overall, since
$\mu \le n$ is the maximum excess). It solves $\rank_1$ queries in time
$\Oh{\log\frac{\mu}{r}}$ if we represent its internal bitvector $H$ of $\Oh{r}$
bits with a structure that solves $\rank$ and $\select$ in constant time
\cite{Cla96}. Note that, since the excess changes by $\pm 1$ across positions,
it changes by $\pm \beta$ across buckets, and thus consecutive elements in the
ladder differ by at most $\beta$. Therefore, it must be $\mu \le r\beta$, and
the time for the $\rank$ operation is $\Oh{\log\beta}=\Oh{\log\log n}$.

\subsection{Range minima and maxima} \label{sec:rmq-large}

If both $i$ and $j$ fall inside the same bucket, then operations $\rmq(i,j)$ 
and relatives are solved inside their bucket.
Otherwise, the minimum might fall in the bucket of $i$, $k_1=\lceil
i/\beta\rceil$, in that of $j$, $k_2=\lceil j/\beta\rceil$,
or in a bucket in between. Using the rmM-trees of buckets $k_1$ and $k_2$, we
find the minimum $\mu_1$ in the range $[i - (k_1-1)\beta,\beta]$ of bucket 
$k_1$, and convert it to a global excess, $\mu_1 \leftarrow \mu_1 + e[k_1-1]$. 
We also find the minimum $\mu_2$ in the range $[1,j - (k_2-1)\beta]$ of 
bucket $k_2$, and convert it to $\mu_2 \leftarrow \mu_2 + e[k_2-1]$. The 
problem is to find the minimum in the intermediate buckets,
$\mu_3 \leftarrow \min_{k_1 < k < k_2} m[k]$. Once we have this, we easily
solve $\rmq(i,j)$ as the position of $\mu_1$ if $\mu_1 \le \min(\mu_3,\mu_2)$,
otherwise as the position of $\mu_3$ if $\mu_3 \le \mu_2$, and otherwise as the
position of $\mu_2$ (recall that we want the leftmost position of the minimum).

In the original work \cite{NS14}, they use the most well-known classical
solution to range minimum queries \cite{BF00}. While it solves the problem
for query $\rmq$, it decomposes the query range $m[k_1+1,k_2-1]$ into 
overlapping subintervals, and thus it cannot be used to solve the other related
queries, such as counting the number of occurrences of the minimum or finding
its $q$th occurrence. As a result, they resort to complex fixes to handle
each of the other related operations in constant time.

If we can allow ourselves to use $\Oh{\log\log n}$ time for the operations,
then a much simpler and elegant solution is possible, using a less known data
structure for range minimum queries \cite{YA10}. It uses $\Oh{n'\log n'}$ 
words, which is $\Oh{n/\log n}$ bits, and solves queries in constant time.
The most relevant feature of this solution is that it reduces the query on
interval $m[k_1+1,k_2-1]$ to disjoint subintervals, which allows solving the
related queries we are interested in.

Assume $n'$ is a power of $2$ and consider a perfect binary tree on top of 
array $m[1,n']$, of height $\lceil \log n'\rceil$. The tree nodes with height
$h$ cover disjoint areas of $m$, of length $2^h$. The tree is stored as a
heap, so we identify the nodes with their position in the heap, starting from
1, and the children of the node at position $v$ are at positions $2v$ 
and $2v+1$.

For each node $v$ covering $m[s,e]$, we store two arrays with the
left-to-right and right-to-left minima in $m[s,e]$, that is, we store
$L[v][p]=\min \{ m[s],\ldots,m[s+p] \}$ and
$R[v][p]=\min \{ m[e-p],\ldots,m[e] \}$ for all $0 \le p \le e-s$.
Their size adds up to $\Oh{n'\log n'}$ cells, or
$\Oh{n'\log n'\log n}=\Oh{n/\log n}$ bits.

Let us call $k=k_1+1$ and $k'=k_2-1$. To find the minimum in $m[k,k']$, we 
compute the lowest node $v$ that covers $[k,k']$. Node $v$ is found as follows:
we compute the highest bit where the numbers $k-1$ and $k'-1$ differ. If this is
the $h$th bit (counting from the left), then node $v$ is of height $h$, and it 
covers the $\ell$th area of $m$ of size $2^h$ (left-to-right), where
$\ell=\lceil k/2^h\rceil$. That is, it holds $v = n'/2^h+\ell-1$ and the range
it covers is $m[s,e] = m[(\ell-1)2^h+1,\ell\, 2^h]$.

The value of $h$ can be computed as $h=\lfloor \log
((k-1)~\mathrm{xor}~(k'-1))\rfloor$\footnote{The $\mathrm{xor}$ operator
        takes two integers and performs the bitwise logical exclusive-or 
operation on them, that is, on each pair of corresponding bits.}. 
If operations $\log$ and $\mathrm{xor}$ are not allowed in the computation model, we can
easily simulate them with small global precomputed tables of size 
$\Oh{\sqrt{n'}}$, 
which can process any sequence of $\log(n')/2$ bits (note that computing $\log$
requires just to find the most significant 1 in the computer word).

Now we have found the lowest node $v$ that covers $[s,e] \supseteq [k,k']$ in
the perfect tree. Therefore, for $p=(s+e-1)/2$, the left child $2v$ of $v$
covers $m[s,p]$ and its right child $2v+1$ covers $m[p+1,e]$. Then, the minimum
of $m[k,k']$ is either that of $m[k,p]$ (which is available at $R[2v][p-k]$) or
that of $m[p+1,k']$ (available at $L[2v+1][k'-p-1]$). We return the minimum of
both. 

This general mechanism is used to solve all the queries related to $\rmq$,
as we see next.

\subsubsection{Solving $\rmq(i,j)$ and $\rMq(i,j)$} \label{sec:rmq}

The only missing piece for solving $\rmq(i,j)$ is to find the leftmost 
position of the minimum in $m[k,k']$. To do this we store other two arrays,
$Lp$ and $Rp$, with the leftmost positions of the minima of the bucket ranges
represented in $L$ and $R$, respectively. That is, if $v$ covers $m[s,e]$, then
$Lp[v][p] = \rmq((s-1)\beta+1,(s+p)\beta)$ and
$Rp[v][p] = \rmq((e-p)\beta+1,e\beta)$.

Thus, once we have the node $v$ that covers $[k,k']$, there are two
choices: If $R[2v][p-k] \le L[2v+1][k'-p-1]$ (i.e., the minimum appears 
in the subrange $m[k,p]$), the leftmost position is $Rp[2v][p-k]$. Otherwise 
(i.e., the minimum appears only in the subrange $m[p+1,k']$) the leftmost
position is $Lp[2v+1][k'-p-1]$. 

Note that any entry from the array $L/R$ can be
obtained on the fly from the corresponding entry of $Lp/Rp$ and the bucket
array $m[]$, hence $L/R$ are only conceptual and we do not store them.
Furthermore, the arrays $Lp/Rp$ are only accessed by nodes that are the
right/left children of their parent, thus we only store one of them in each
node. 

Operation $\rMq(i,j)$ is solved analogously (needing similar structures $R$, 
$L$, $Lp$ and $Rp$ regarding the maxima).

\subsubsection{Solving $\mincount(i,j)$} \label{sec:mincount}

To count the number of times the minimum appears, we first compute 
$\mu=\min(\mu_1,\mu_2,\mu_3)$, and then add up its occurrences in each of the
three ranges: we add up $\mincount(i - (k_1-1)\beta,\beta)$ in bucket $k_1$ if
$\mu=\mu_1$, $\mincount(1,j - (k_2-1)\beta)$ in bucket $k_2$ if $\mu=\mu_2$,
and the number of times the minimum appears in $[(k-1)\beta+1,k'\beta]$ 
(i.e., inside buckets $k$ to $k'$) if $\mu=\mu_3$. To compute this last number,
we store two new arrays, $Ln$ and $Rn$, giving the number of times the minimum
occurs in the corresponding areas of $L$ and $R$, that is,
$Ln[v][p] = \mincount((s-1)\beta+1,(s+p)\beta)$ and
$Rn[v][p] = \mincount((e-p)\beta+1,e\beta)$. 

Thus, if $R[2v][p-k] < L[2v+1][k'-p-1]$, then the minimum appears only on the
left, and the count in buckets $k$ to $k'$ is $Rn[2v][p-k]$. If 
$R[2v][p-k] > L[2v+1][k'-p-1]$, it appears only on the right, and the count is
$Ln[2v+1][k'-p-1]$. Otherwise, it appears in both and the count is
$Rn[2v][p-k] + Ln[2v+1][k'-p-1]$.  Once again, a node needs to store only
$Ln$ or $Rn$, not both.

\subsubsection{Solving $\minselect(i,j,q)$}

To solve $\minselect(i,j,q)$ we must see if $q$ falls in the bucket of $k_1$,
in the bucket of $k_2$, or in between. We start by considering $k_1$, if
$\mu=\mu_1$. In this case, we compute $q_1 = \mincount(i - (k_1-1)\beta,\beta)$,
the number of times $\mu$ occurs inside bucket $k_1$. If $q \le q_1$, then the
$q$th occurrence is inside it, and we answer
$\minselect(i - (k_1-1)\beta,\beta,q)$. If $q>q_1$, then we continue, with
$q \leftarrow q-q_1$.

If $\mu$ appears between $k_1$ and $k_2$, that is, if $\mu=\mu_3$, we compute
$q_3 = \mincount((k-1)\beta+1,k'\beta)$ as in Section~\ref{sec:mincount}. Again,
if $q \le q_3$, the answer is the $q$th occurrence of the
minimum in buckets $k$ to $k'$. If $q > q_3$, we just set $q \leftarrow q-q_3$.
Finally, if we have not yet solved the query, we return
$\minselect(1,j - (k_2-1)\beta,q)$ within bucket $k_2$.

To find the $q$th occurrence of $\mu$ in the buckets $k$ to $k'$, we make use
of the arrays $Ln$ and $Rn$. If $\mu < R[2v][p-k]$, then the answer is to be
found in the buckets $p+1$ to $k'$. If, instead, $\mu = R[2v][p-k]$, then there
are $Rn[2v][p-k]$ occurrences of $\mu$ in $m[k,p]$. Thus, if $q\le Rn[2v][p-k]$,
we must find the $q$th occurrence of $\mu$ in buckets $k$ to $p$. If instead
$q > Rn[2v][p-k]$, we set $q \leftarrow q- Rn[2v][p-k]$ and find the $q$th
occurrence of the minimum in buckets $p+1$ to $k'$.

Let us find the $q$th occurrence of $\mu$ in buckets $k$ to $p$ (the other case
is symmetric, using $L$ instead of $R$). The minimum in $m[k,p]$ is $\mu$. It
also holds that the minimum in $m[k+l,p]$ is $\mu$, for all $0 \le l \le g$,
for some number $g \ge 0$, and then $m[k+g+1,p] > \mu$. Those intervals are
represented in the cells $R[2v][p-k]$ to $R[2v][p-k-g]$, and the number of times
$\mu$ occurs in them is in $Rn[2v][p-k]$ to $Rn[2v][p-k-g]$. Therefore, our 
search for the $q$th minimum spans a contiguous area of $Rn[2v]$: we want
to find the largest $l \ge 0$ such that $Rn[2v][p-k-l] \ge q$. This means
that the $q$th occurrence of $\mu$ in buckets $k$ to $p$ is in bucket
$k+l$, in whose rmM-tree we must return $\minselect(1,\beta,Rn[2v][p-k-l]-q+1)$.

To find $l$ fast, we record all the values $Rn[2v][\cdot]$ in 
complemented unary (i.e., number $x\ge 0$ as $0^x1$) in a bitvector $C$. Then,
each $0$ counts an occurrence of the minimum and each $1$ counts a bucket. 
To find $l$, we compute
$y=\select_1(C,p-k)-(p-k)$, the sum of the values up to $Rn[2v][p-k]$, and then
$l'=\select_0(C,y-q+1)-(y-q)$ is the desired cell $Rn[2v][p-k-l]$, thus 
$l=p-k-l'$.

We use again the \emph{sarray} bitvector of Okanohara and Sadakane \cite{OS07}.
It solves $\select_1$ in constant time and $\select_0$ in the same time as
$\rank$.  There is a 1 per cell in $Rn$, so the global space is at most $(n'\log
n + \Oh{n'})\log n' = \Oh{n/\log n}$ bits.  Since the distance between
consecutive 1s is at most $\beta$, the time to compute $\select_0$ is
$\Oh{\log\beta}=\Oh{\log\log n}$.

Note, in passing, that bitvector $C$ can replace $Rn[2v]$, as it can compute
any cell $Rn[2v][x]=\select_1(C,x)-\select_1(C,x-1)-1$ in constant time.
Therefore we can use those bitvectors instead of storing arrays $Rn$ and $Ln$,
thus avoiding to increase the space further.

\subsection{Rank and select operations} \label{sec:rs}

The various basic and extended $\rank_x$ and $\select_x$ operations are 
implemented similarly as the more complex operations. For $\rank_x$, we 
store the $\rank_x$ value at the beginning of each bucket, in an array 
$r_x[1,n']$, and then compute $\rank_x(i)=r_x[k]+\rank_x(i-(k-1)\beta)$ inside 
the rmM-tree of bucket $k = \lceil i/\beta \rceil$. For $\select_x(j)$,
we store the $r_x[k]$ values in a bitvector $B_x[1,n]$ with 
$B_x[k+r_x[k]]=1$ for all $k$, then the bucket $k$ where the answer lies is 
$k=\select_0(B_x,j)-j+1$, inside whose rmM-tree we must solve 
$\select_x(j-r_x[k])$. Again, with the bitvectors of Okanohara and Sadakane
\cite{OS07}, we do not need to store $r_x$ because its cells are computed in
constant time as $r_x[k] = \select_1(B_x,k)-\select_1(B_x,k-1)-1$, the space
used is $n'\log n+\Oh{n'} = \Oh{n/\log^2 n}$ bits, and the time to compute
$\select_0$ is $\Oh{\log\log n}$ because there are at most $\beta$ 0s per 1 in
$B_x$.

\section{Implementation and Experimental Results}

We now describe an engineered implementation based on our theoretical 
description, and experimentally evaluate it. Engineered implementations often
replace solutions with guaranteed asymptotic complexity by simpler variants 
that perform better in most practical cases. Our new theoretical version is
much simpler than the original \cite{NS14}, and thus most of it can be
implemented verbatim. Still, we further simplify some parts to speed them up
in practice. As a result, our implementation does not fully guarantee
$O(\lg\lg n)$ time complexity, but it turns out to be faster than the 
state-of-the-art implementation that uses $O(\log n)$ time. As this latter
implementation essentially uses one binary rmM-tree for the whole sequence, our
experiments show that our new way to handle inter-bucket queries is useful in 
practice, reducing both space and time.

\subsection{Implementation}

We use a fixed bucket size of $\beta=2^{15}$ parentheses (i.e., $4$KB).
Since the \emph{relative} excess inside each bucket are in the range 
$[-2^{15}, 2^{15}]$ the fields of the nodes of each rmM-tree are stored 
using $16$-bit integers. To reduce space, we get rid of the $v.e$ fields by 
storing $v.m$ and $v.M$ in \emph{absolute} form, not relative to 
their rmM-subtree.\footnote{These values are absolute within their current 
bucket; they are still relative to the beginning of the bucket (otherwise they 
would not fit in $16$ bits).} This is because the field $v.e$ is used only to 
convert relative values to
absolute.\footnote{Instead, relative values allow making the structure
dynamic, as efficient insertions/deletions become possible \cite{NS14}.}
This reduces the space required by the rmM-tree nodes from 8
to 6 bytes (or 4 bytes if the field \emph{v.n} is not required, as it is
used only in the more complicated operations). The block size of each
rmM-tree, $b$, is parameterized and provides a space-time tradeoff:  the bigger
the block size, the more expensive it is to perform a full scan.  The sequential
scan of a block is performed by lookup tables that handle \emph{chunks} of
either $8$ or $16$ bits. Preliminary tests yielded the following values to be 
reasonable for $b$: $512$ bits (with lookup tables of $8$ bits) and
$1024/2048$ bits (with lookup tables of $16$ bits).  In particular, for $b=1024$
our rmM-trees have height $h = \lg (\beta/b) = 5$ and a sequential scan of a
block requires up to $64$ table lookups.

The \emph{bucket} arrays $e[], m[]$ and $M[]$ are stored in heap form, as 
described. The special tree $T'$ of Section~\ref{sec:fwd-large} is built
using a stack-based folklore algorithm that finds the previous-smaller-value
of each element in array $m[]$ in linear time and space (that is, $\Oh{n/\beta}$
words).  The \emph{ladder} decomposition and pointers to ancestors at
distances $2^k$ (for some $k$) in $T'$ are implemented verbatim.  To find the
target bucket for operation $\fwdsearch$ we sequentially iterate over
$k=0,1,\ldots$ to find an ancestor whose minimum excess is lower than the
target, then we perform a sequential search in its ladder to find the target
bucket.  Although this implementation does not guarantee $\Oh{\log\log n}$
worst case time, it is cache-friendly and faster than doing a binary search over
the list of sampled ancestors or using the \emph{sarray} bitmap representation
to accelerate the search. On the real datasets that were used for
the experiments, the height of $T'$ was in all cases less than $10$, which
fully justifies a sequential scan. A more
sophisticated implementation could resort to the guaranteed $O(\lg\lg n)$-time
method when it detects that the ladder or the list of ancestors are long enough.

For operation $\rmq(i, j)$ and relatives, the perfect binary tree of 
Section~\ref{sec:rmq-large} is implemented verbatim, except that the bitvector
$C$ is not implemented; a sequential search in $Rn/Ln$ is carried out instead
for $\minselect$. The extended $\rank$ and $\select$ operations were not yet
implemented.

\subsection{Experiemental setup}

To measure the performance of our new implementation we used two public
datasets\footnote{Available at
\url{http://www.inf.udec.cl/~josefuentes/sea2015/}}: \texttt{wiki}, the XML tree
topology of a Wikipedia dump with $498,753,916$ parentheses and \texttt{prot},
the topology of the suffix tree of the Protein corpus from the Pizza\&Chili
repository\footnote{Available at \url{http://pizzachili.dcc.uchile.cl/}} with
$670,721,008$ parentheses. 

We replicate the benchmark methodology used by
Arroyuelo et al.~\cite{ACNS10}: we fix a probability $p \in [0, 1]$ and generate a
\emph{sample dataset} of nodes by performing a depth-first traversal of the tree
where we descend to a random child and also descend to each other child
with probability $p$.  All datasets generated consist of at least
$200,000$ nodes.  Setting $p=0$ emulates random root-to-leaf
paths while $p=1$ provides a full traversal of the tree. Intermediate
values of $p$ emulate other tree traversals that occur, for example,
when solving XPath queries or performing approximate string matching on suffix 
trees. We benchmark the operations $\open/\close/\enclose$ for $p=0.00$,
$0.25$, and $0.50$. We also benchmark operation $\rmq(i,j)$ by choosing 
200,000 pairs $i<j$ at random and classifying the results according to $j-i$.

All the experiments were ran on a Intel(R) Core(TM) i5 running at $2.7$GHz with
$8$GB of RAM running Mac OS X 10.10.5.  Our implementation is single-threaded,
written in \texttt{C++}, and compiled with \texttt{clang} version
$7.0.0$ with the flags \texttt{-O3} and \texttt{-DNDEBUG}.

As a baseline we use the \texttt{C++} implementation available in the
Succinct Data Structures Library \footnote{Available at
\url{github.com/simongog/sdsl-lite}}(SDSL), which provides an $\Oh{\log n}$-time
implementation based on the description of Arroyuelo et al.~\cite{ACNS10}.
This library is known for its excellent implementation quality. In particular,
this implementation also stores the fields $v.m$ and $v.M$ in absolute form
and discards $v.e$. It also does not store $v.n$, as it does not implement the
more complex operations associated with it. For this reason, we will only
compare the structures on the most basic primitives 
$\open/\close/\enclose/\rmq$ that are also implemented in SDSL. Also, for 
fairness, we {\em do not} account for the space of the field $v.n$ in our
structure.

\subsection{Experimental results}

\begin{figure}[t!]
\centering
\begin{tabular}{cc}
\subf{\includegraphics[width=60mm]{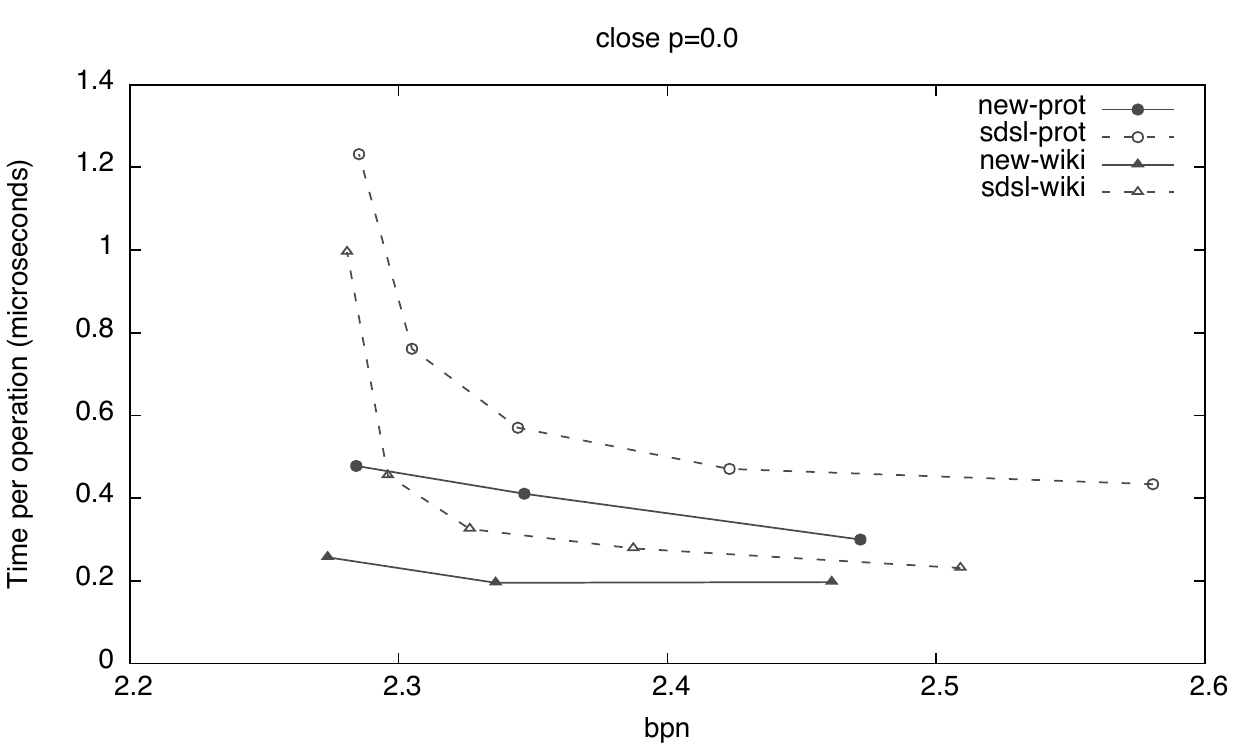}} &
\subf{\includegraphics[width=60mm]{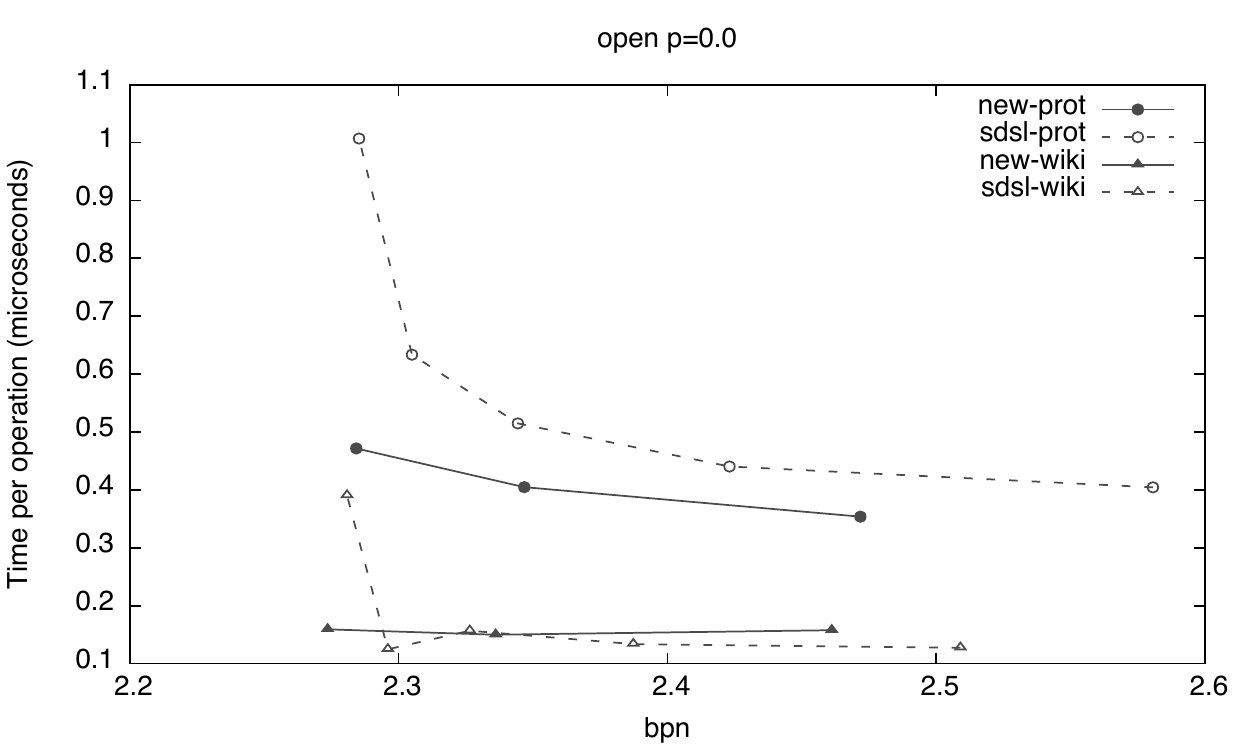}} \\
\subf{\includegraphics[width=60mm]{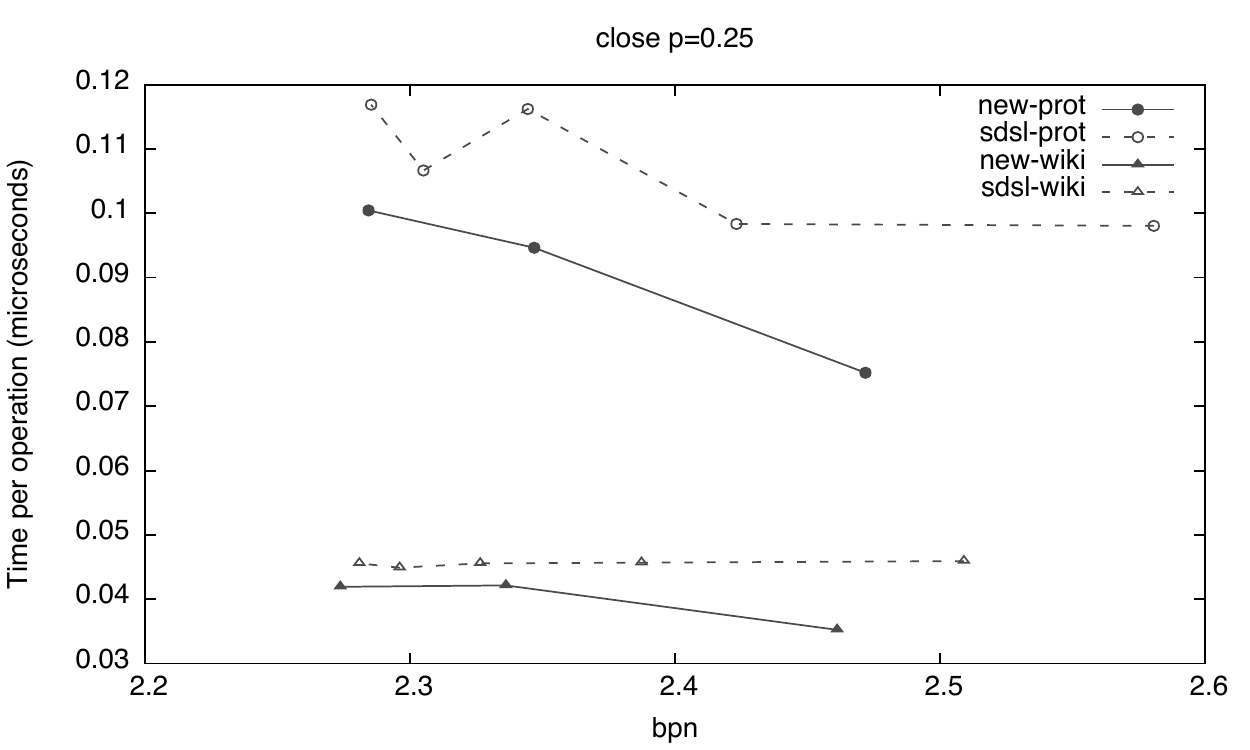}} &
\subf{\includegraphics[width=60mm]{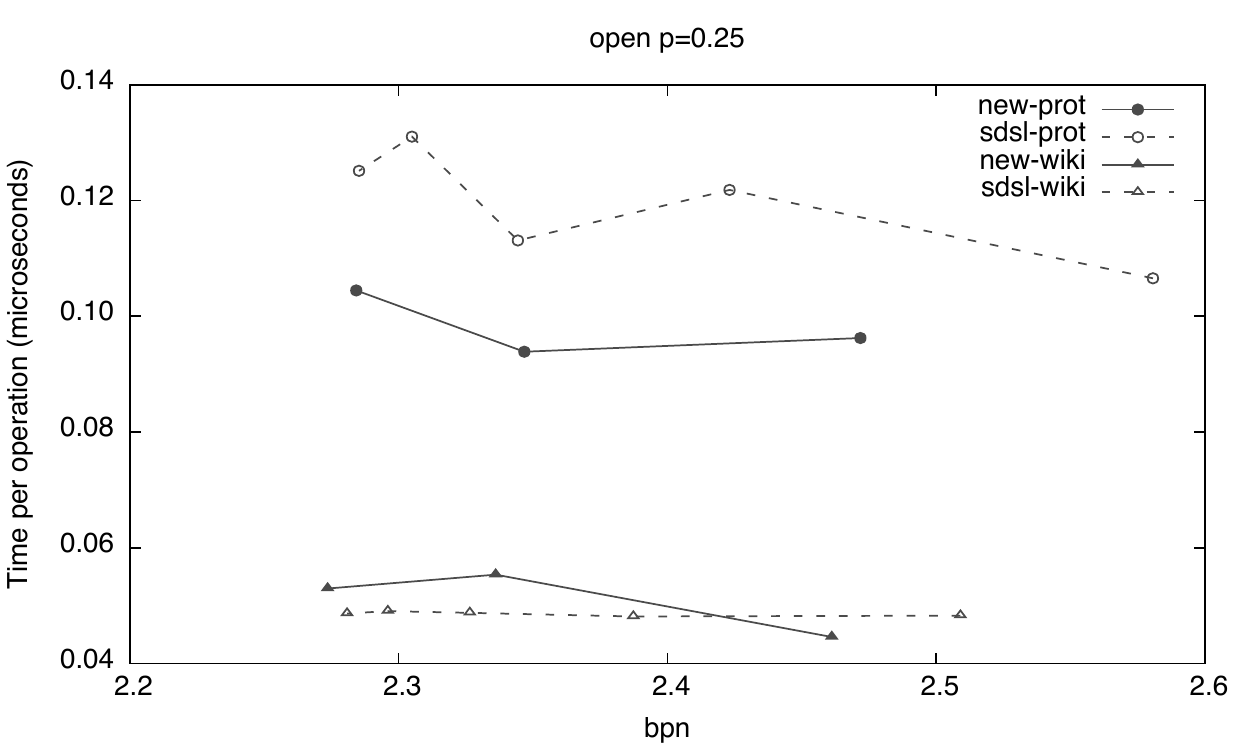}} \\
\subf{\includegraphics[width=60mm]{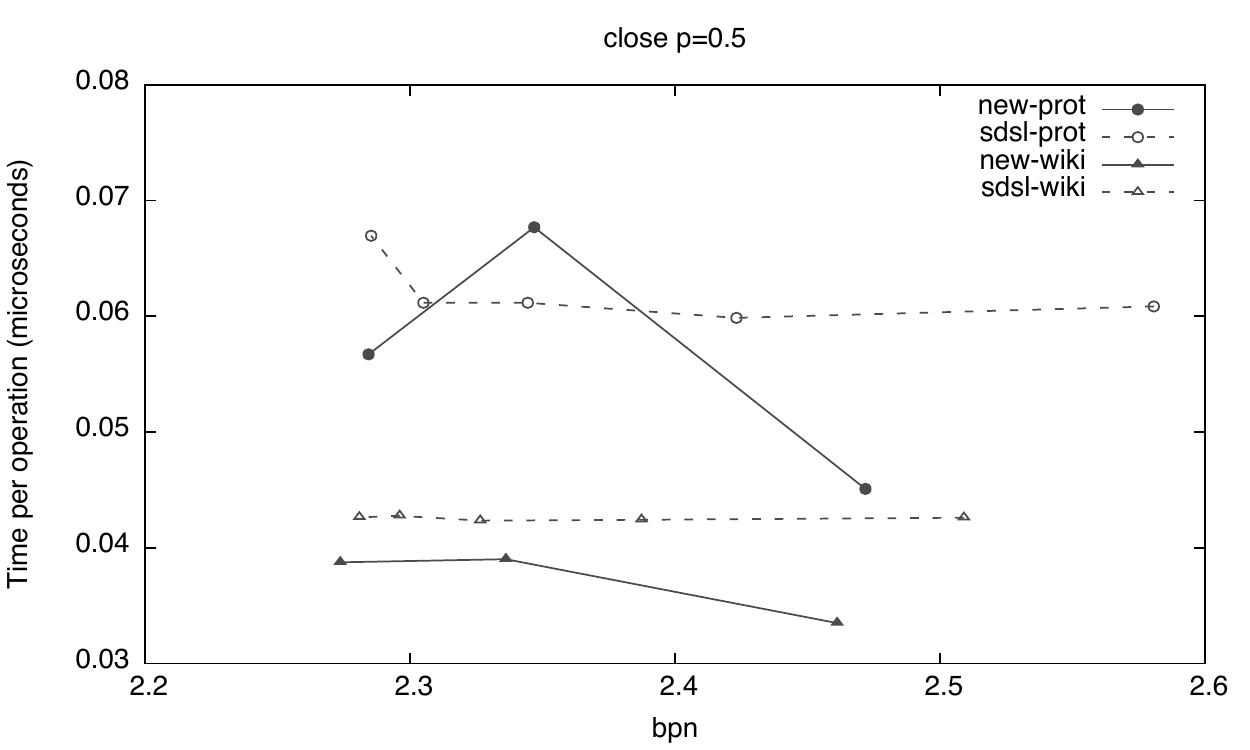}} &
\subf{\includegraphics[width=60mm]{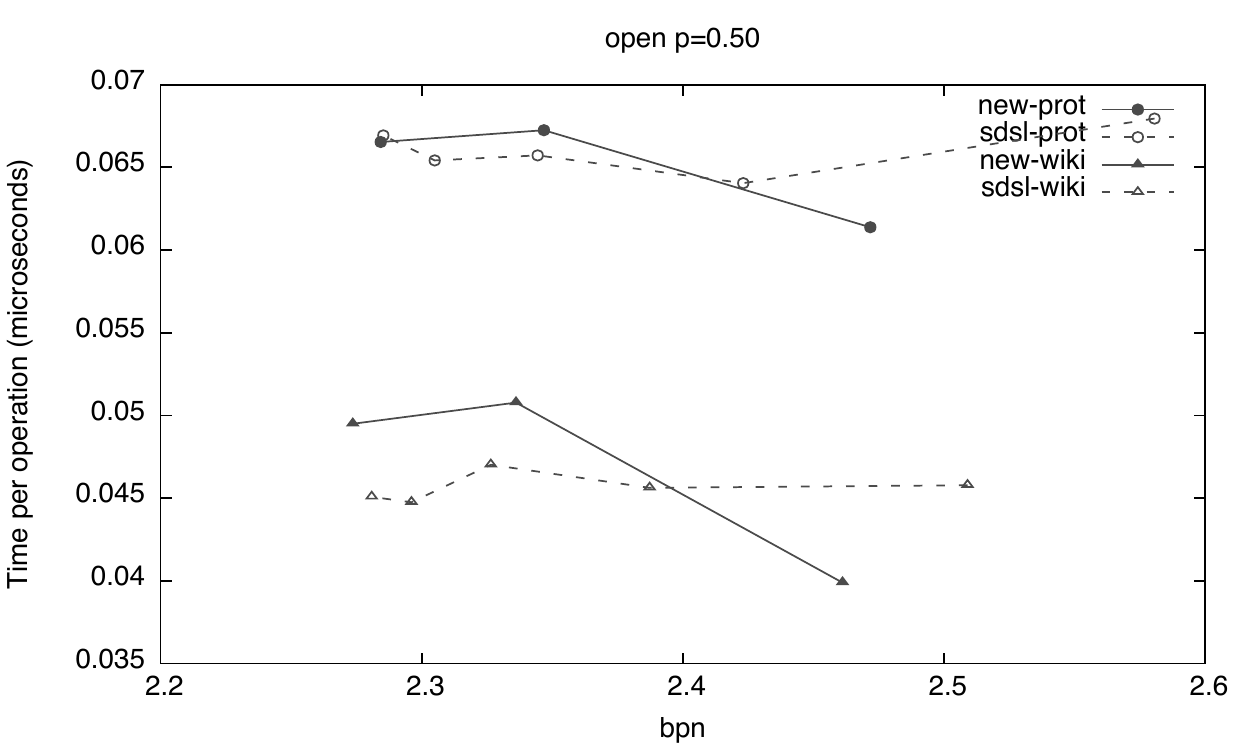}} 
\end{tabular}
\caption{Space-time tradeoffs for our new implementation and the SDSL baseline,
for operations $\close$ (left) and $\open$ (right).}
\label{fig:data}
\end{figure}

\begin{figure}[t!]
\centering
\begin{tabular}{cc}
\subf{\includegraphics[width=60mm]{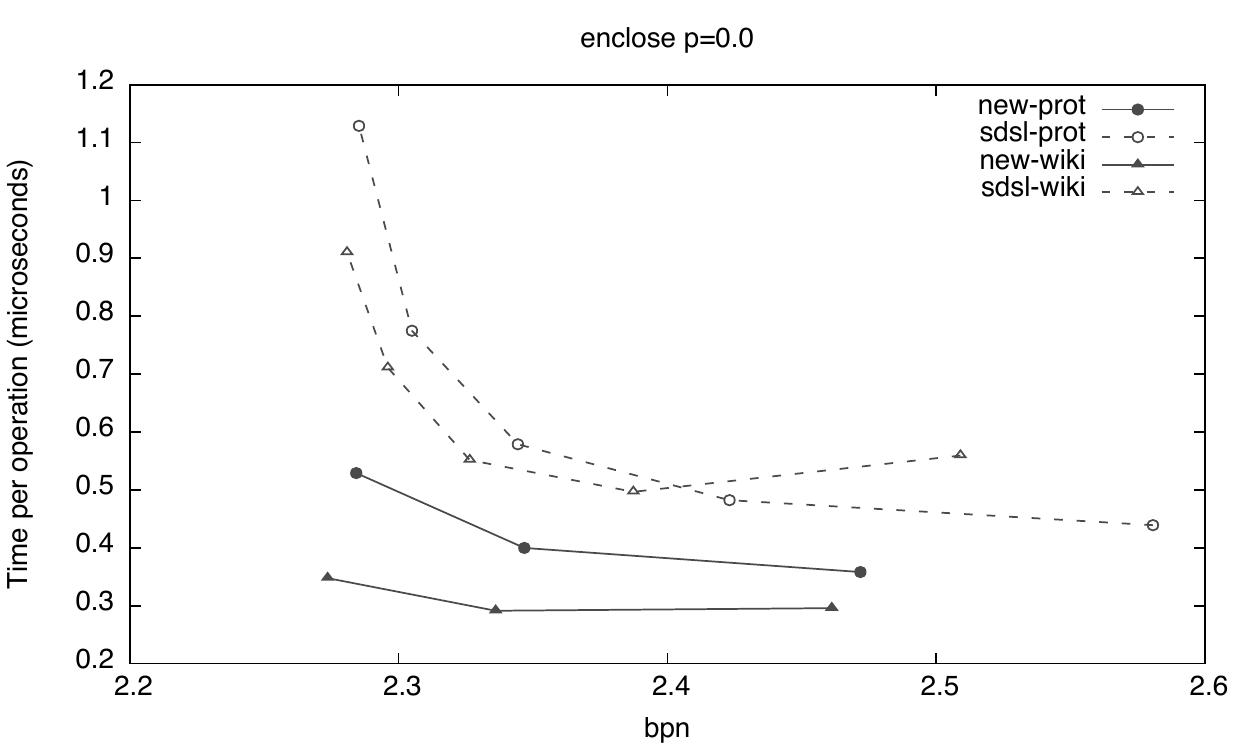}} &
\subf{\includegraphics[width=60mm]{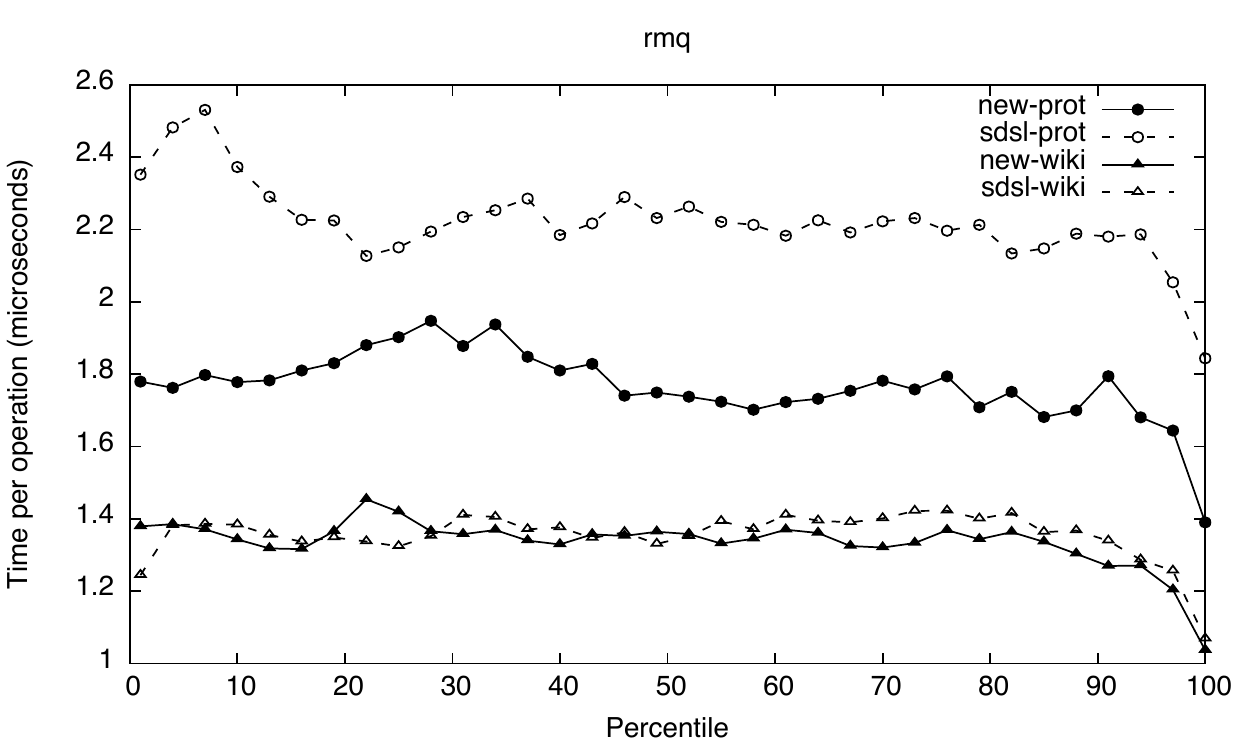}} \\
\subf{\includegraphics[width=60mm]{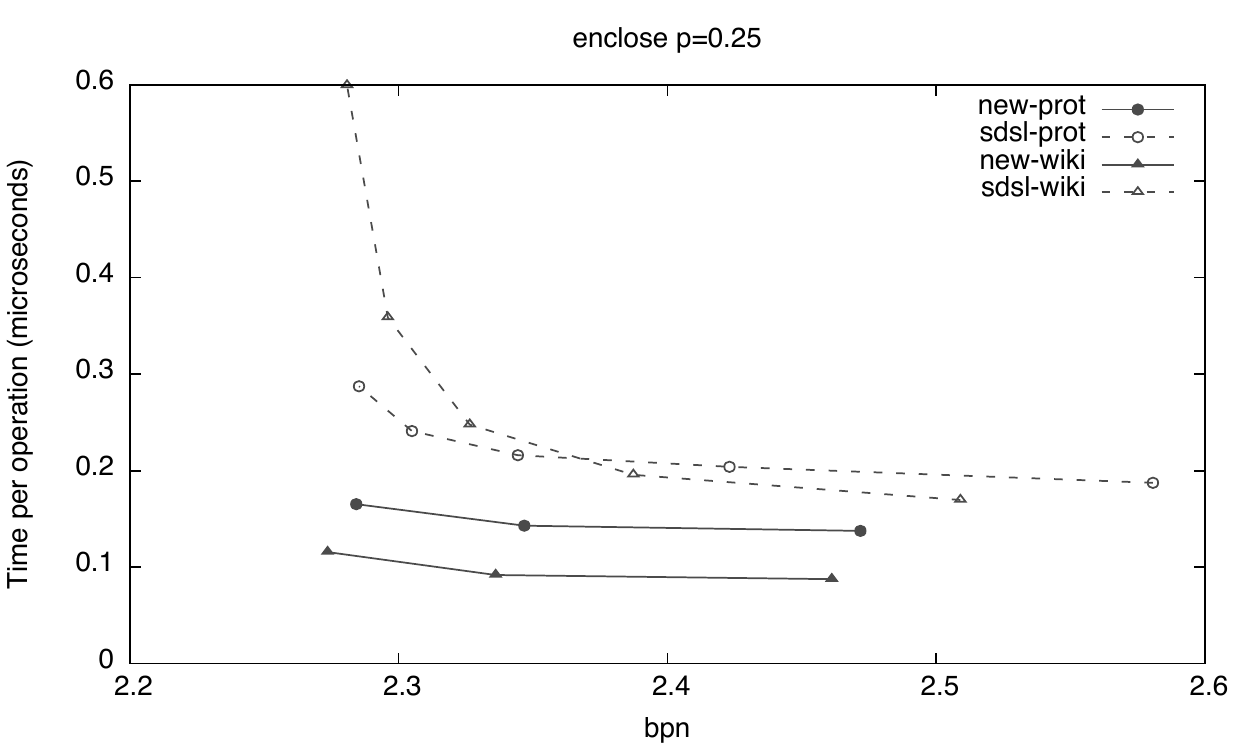}} &
\subf{\includegraphics[width=60mm]{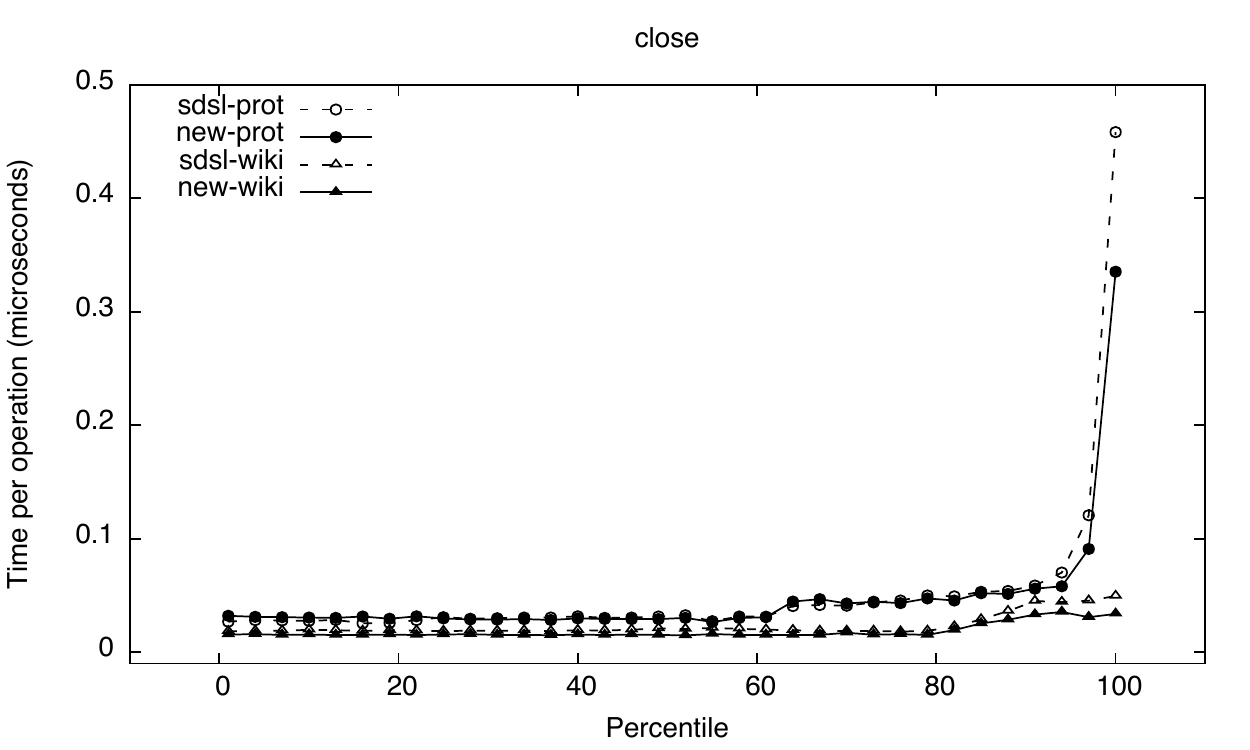}} \\
\subf{\includegraphics[width=60mm]{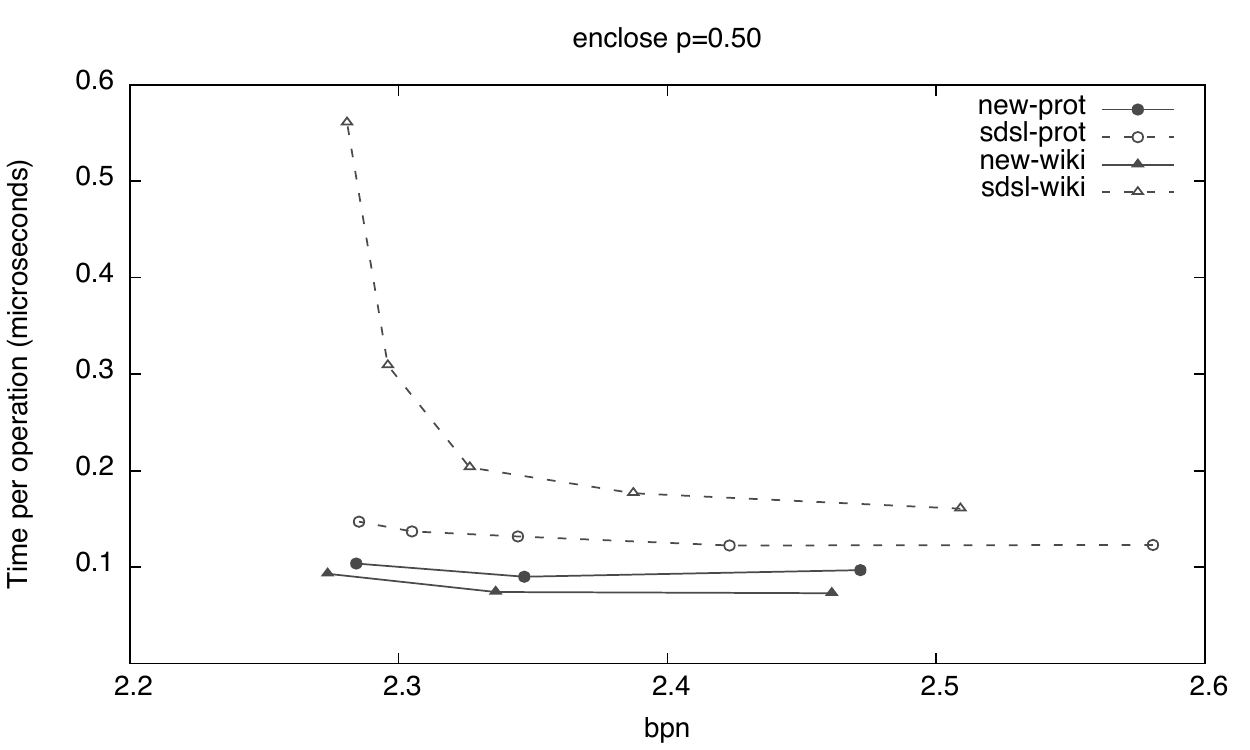}} &
\subf{\includegraphics[width=60mm]{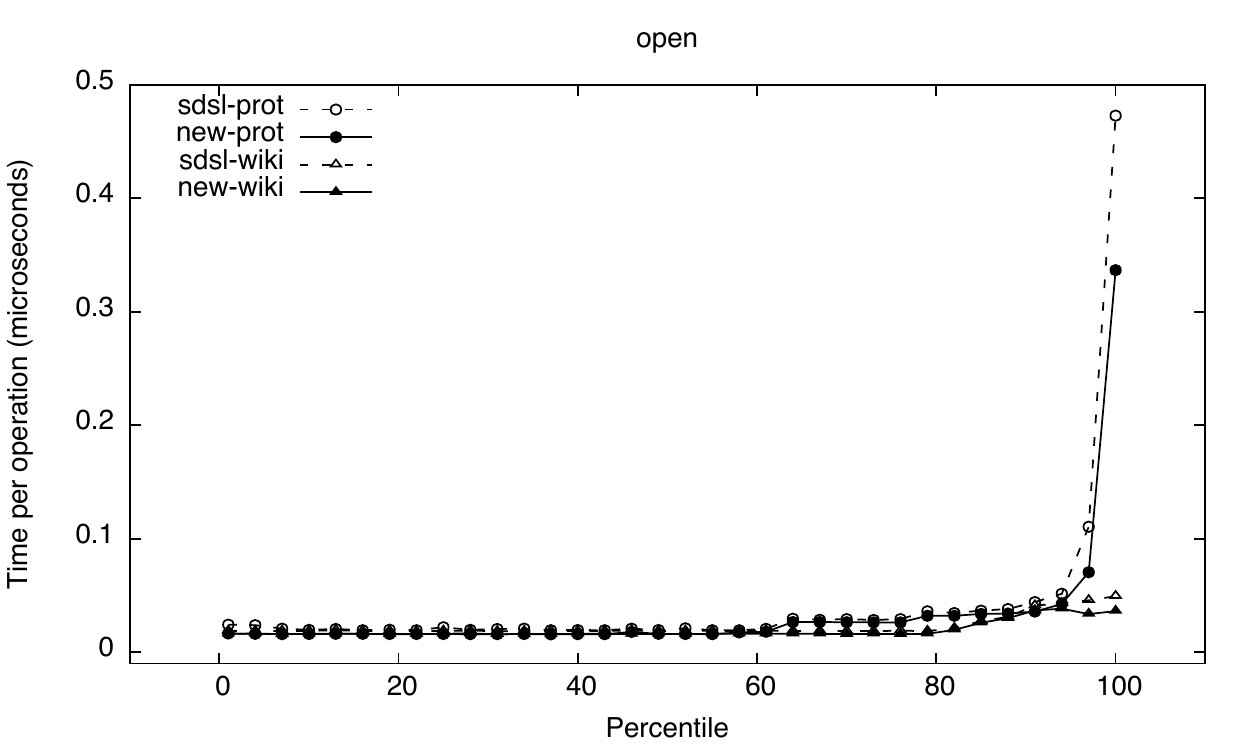}} \\
\end{tabular}
\caption{Space-time tradeoffs for our new implementation and the SDSL baseline,
for operation $\enclose$ (left). On the right, the results as a function
of the distance traversed in the parenthesis sequence for $\rmq$, $\close$, and 
$\open$.}
\label{fig:data2}
\end{figure}

Figures~\ref{fig:data} and \ref{fig:data2} (left) show the results for 
$\open/\close/\enclose$ operations with
different values of $p$.  The times reported are in microseconds and are the
average obtained by performing the operation over all the nodes of a
dataset generated for a given parameter value $p$. 
The space is reported in bits per node (\emph{bpn}).
The \emph{new-} prefix refers
to the implementation of our new structure, while \emph{sdsl-} refers
to the SDSL implementation. The three space-time tradeoffs shown in our new
implementation correspond to $b=512$, $1024$, and $2048$ (a larger $b$ obtains
lower space and higher time).

For operation $\close$, our implementation is considerably faster than
SDSL, while using essentially the same space. For $p=0.0$, we are up to $4$
times faster when using the least space. For larger $p$, the operations
becomes much faster due to the locality of the traversals, and the time
differences decrease, but it they are still over 10\%.

Our implementation is still generally faster for $\open$ on \texttt{prot},
whereas on \texttt{wiki} SDSL takes over for larger $p$ values. The maximum
advantage in our favor is seen on operation $\enclose$, where our implementation
is $2$--$6$ times faster when using the least space, with the only exception of
\texttt{prot} with $p=0.50$, where we are only 30\% faster.

For operation $\rmq$ we show the results classified by $j-i$, cut into 100 
percentiles. Figure~\ref{fig:data2} (top right) shows the results.
Both structures use the same space, about $2.34$ bits per node.
On \texttt{prot} we are significantly faster in almost all the spectrum, while 
on \texttt{wiki} we are generally faster by a small margin. The difference owes
to the fact that the tree of \texttt{prot} is much deeper, and therefore the
traversals towards the $\rmq$ positions are more random and less cache-friendly.
In \texttt{wiki}, the root and the highest nodes are the answers to random
$\rmq$s in most cases, so their rmM-trees are likely to be in cache from 
previous queries. On the other hand, we note that the times are basically
constant as a function of $j-i$.

The other plots on the right of Figure~\ref{fig:data2} we show how the times for
operation $\close$ and $\open$ evolve as a function of the difference between 
the position that is queried and the one where the answer is found. We use the 
configuration with about $2.34$ bits per node for both implementations, and
average the query times over all the tree nodes. In general, only a slight 
increase is observed as the distance grows. In the larger sequence {\tt prot}, 
however, there is a sharp increase for the largest distances. This is not 
because the number of operations grows sharply, but it rather owes to a 10X 
increase in the number of cache misses: traversing the
longest distances requires accessing various rmM-tree nodes that no longer fit
in the cache. Note that the highest times, around 0.5 $\mu$s, are indeed the
typical times obtained in Figure~\ref{fig:data} with $p=0.0$, where most of 
the nodes traversed produce cache misses.

\section{Conclusions}

We have described an alternative solution for representing ordinal trees of $n$
nodes within $2n+\Oh{n/\log n}$ bits of space, which solves a large number of
queries in time $\Oh{\log\log n}$. While the original solution upon which we
build \cite{NS14} obtains constant times, it is hard to implement and only
variants using $\Oh{\log n}$ time had been successfully implemented.  We have
presented a practical implementation of our solution and have experimentally
shown that, on real hundred-million node trees, it achieves better space-time
tradeoffs than current state-of-the-art implementations. This shows that the
new design has not only theoretical, but also practical value. Our new
implementation is publicly available at 
{\tt www.dcc.uchile.cl/gnavarro/software}.

\bibliographystyle{plain}
\bibliography{paper}

\end{document}